\documentclass[twocolumn,showpacs,amsmath,amssymb,aps,groupedaddress]{revtex4}
\usepackage[latin9]{inputenc}
\usepackage{graphicx}
\usepackage{amsmath}
\usepackage{color}
\usepackage{epsfig}

\begin{document}

\title{The linear spectrum of a quantum dot coupled to a nano-cavity}

\author{G. Tarel}
\affiliation{Institut de Théorie des Phenomènes Physiques, Ecole Polytechnique Fédérale de Lausanne, CH-1015 Lausanne EPFL, Switzerland}
\author{V. Savona}
\affiliation{Institut de Théorie des Phenomènes Physiques, Ecole Polytechnique Fédérale de Lausanne, CH-1015 Lausanne EPFL, Switzerland}

\date{\today}

\begin{abstract}
We develop a theoretical formalism to model the linear spectrum of a quantum dot embedded in a high quality cavity, in presence of an arbitrary mechanism modifying the homogeneous spectrum of the quantum dot. Within the simple assumption of lorentzian broadening, we show how the known predictions of cavity quantum electrodynamics are recovered. We then apply our model to the case where the quantum dot interacts with an acoustic-phonon reservoir, producing phonon sidebands in the response of the bare dot. In this case, we show that the sidebands can sustain the spectral response of the cavity-like peak even at moderate dot-cavity detuning, thus supporting recent experimental findings.
\end{abstract}


\maketitle

\section{Introduction}

The description of semiconductor quantum dots (QDs) as isolated atomic-like quantum systems is largely oversimplified.
The confined electrons and holes in a QD, in fact, interact rather efficiently with both the electronic and vibrational
\cite{vasanelli, borri, krumm,muljarov,Besombes} degrees of freedom of the semiconductor environment, in ways that can only
be described beyond the simple perturbation theory. In addition to the semiconductor
medium, QDs interact with the surrounding electromagnetic field, especially if embedded in a photonic structure with sharp
electromagnetic resonances. As an example, the electromagnetic field can vehiculate an excitation transfer between
two distant QDs with non-overlapping electronic states \cite{parascandolo, tarel2008,ForchelNatPhys}. If the QD is embedded in a high-quality cavity,
the 3D confinement of electromagnetic field can lead to observation of the strong coupling between one QD and the resonant mode of the
electromagnetic field \cite{Thon,reithmaier, yoshie2004, peter,Hennessy}. This system, however, can not be seen as a perfect parallel to the atom-cavity coupling in cavity quantum electrodynamics (CQED) \cite{Raimond}. The semiconductor
environment, in particular, can affect the system in several ways that have no analogous in its atom-cavity counterpart. A first effect is brought by the coupling of the QD to an external reservoir (e.g. phonons) that can produce a significant change in the homogeneous spectral signature of the dot. As examples we quote the broad sidebands originating by the coupling to longitudinal acoustic phonons beyond perturbation theory \cite{borri,zimmermann,Besombes,Favero,krumm,MildePRB}, or the similar effect due to optical phonons \cite{Stauber,Muljarov2007}. In addition to these homogeneous modifications of the bare QD spectrum, other significant spectral changes can arise when the QD is multiply excited, due to transitions between continuum states in the wetting layer above the QD confining barrier. This effect occurs already at moderate excitation and leads to a sizeable enhancement of off-resonance light emission \cite{Winger2009,Kaniber08,Fiore2009}.

Here, we address the first class of these semiconductor-related effects, where a homogeneous change in the spectrum of the single QD is present. We restrict to the linear spectral response of the cavity-QD system, holding at low excitation density. Using a Maxwell formalism, we show how this effect translates in the overall spectral signature of the QD-cavity system, at varying detuning and QD parameters. Our model accounts for the specific shape of the resonant cavity mode as well as for the microscopic parameters of the QD excitonic transition.

We first develop the general formalism, assuming an arbitrary energy-dependent self-energy for the QD excitonic transition. Then we apply the model to two cases. The first is that of a lorentz-shaped QD line, characterized by a constant broadening $\gamma_0$. In this case, we recover the result that is well known from CQED \cite{Khitrova}. Then, we assume a QD spectrum arising from the coupling to longitudinal acoustic phonons, with the corresponding self-energy modeled within a second-Born approximation \cite{krumm}. The coupling produces phonon sidebands in the bare QD spectral response. We show how the spectral signature of these sidebands is enhanced by the presence of the cavity resonance, even for moderate QD-cavity detuning. The persistence of light emission at the cavity-like peak has been the object of several experimental investigations recently \cite{Kaniber08,Hennessy,Fiore2009,suffczynski2009,Winger2009}. At small detuning, the acoustic phonon mechanism is expected to contribute significantly to this effect, as the energy width of the sidebands is determined by the exciton spatial confinement \cite{zimmermann}, and amounts to 1-2 meV in typical samples. This effect has been recently addressed using phenomenological dephasing to account for the modified QD spectral signature \cite{Naesby,Auffeves09,Cui}, or within a more microscopic approach to the phonon sideband mechanism at zero QD-cavity detuning \cite{MildePRB}. The importance of our work lies in the fact that a general homogeneous mechanism acting on the QD spectrum is modeled, and an explicit expression for the total emission spectrum, accounting for spatial and spectral cavity form factors, is derived.

In Section II, we present the general theoretical formalism. Section III is devoted to deriving the simple CQED result in the limit of lorentz QD broadening. In Section IV, we study the case of a QD coupled to a reservoir of longitudinal acoustic phonons, and discuss how phonon sidebands enhance the cavity-like emission spectrum at finite cavity QD detuning. In Section V, we present our conclusions.

\section{Theoretical formalism}
\subsection{Maxwell equations}

We consider the system of one QD embedded in a resonant nanocavity. The cavity can be of any kind (e.g. pillar \cite{reithmaier},
photonic crystal defect\cite{yoshie2004}, microdiscs\cite{peter}, etc.), with the only assumption that one well distinct resonant
mode exists in the vicinity of the QD transition wavelength. Our objective is to derive the physical parameters characterizing an
effective coupling to this mode, at frequency $\omega_c$, from the microscopic details of the electromagnetic field in the structure.
We assume a QD lying at position ${\bf r_0}$. Typically, this position is selected to lie where the electric
field has maximum amplitude, in order to maximize QD-cavity coupling. In the limit of low QD excitation, the spectra are determined by the linear optical response, and are described by
Maxwell equations for the electric field $\bm{\mathcal{E}}$ coupled to the linear susceptibility tensor of the QD.
Under this assumption,
the steps leading to a set of coupled mode equations are formally the same as in our previous works \cite{parascandolo, tarel2008}. In particular
Maxwell equations are cast into an integral Dyson equation \cite{martin1998}. We denote with $\epsilon=\epsilon({\bm{r}})$
the spatially dependent dielectric constant that characterizes the resonant photonic structure.
In the frequency domain, we have (assuming non magnetic medium and no free charges) :
\begin{multline} \label{Maxwell}
\bm{\nabla}\wedge\bm{\nabla}\wedge
{\bm{\mathcal E}}\left({\bm{r}},\omega\right)-\frac{\omega^2}{c^2}
\bigg[\epsilon({\bm{r}}) {\bm{\mathcal E}}\left({\bm{r}},\omega\right)
\\
\left.+4\pi\int d{\bm{r}}^\prime
\hat{\bm{\chi}}_{QD}\left({\bm{r}},{\bm{r}}^\prime,\omega\right)
\cdot {\bm{\mathcal E}}\left({\bm{r}}^\prime,\omega\right)\right]=0
\,,\nonumber
\end{multline}
where ${\bm{\mathcal E}}\left({\bm{r}},\omega\right)$ is the electric field, $\bm{r}$ is the 3-D position vector, and $\hat{\bm{\chi}}_{QD}$ the $3\times3$ linear optical susceptibility tensor of the QD subsystem. In order to define a hermitic problem, we adopt the standard replacement \cite{SakodaBook}
\begin{equation} \label{Qdef}
 \bm{\mathcal Q}\left({\bm{r}},\omega\right)=\sqrt{\epsilon({\bm{r}})}\bm{\mathcal E}(\bm{r},\omega)\,,
\end{equation}
and
\begin{equation} \label{Upsilon}
\bm{\Upsilon}=\frac{1}{\sqrt{\epsilon({\bm{r}})}}\bm{\nabla}\wedge \{ \bm{\nabla}\wedge \frac{1}{\sqrt{\epsilon({\bm{r}})} } \}\,.
\end{equation}
This leads to the following hermitic problem:
\begin{multline} \label{Qprob}
\bm{\Upsilon}
{\bm{\mathcal Q}}\left({\bm{r}},\omega\right)-\frac{\omega^2}{c^2}
{\bm{\mathcal Q}}\left({\bm{r}},\omega\right)=
\\
+\frac{4\pi\omega^2}{c^2\sqrt{\epsilon({\bm{r}})}}\int d{\bm{r}}^\prime
\hat{\bm{\chi}}_{QD}\left({\bm{r}},{\bm{r}}^\prime,\omega\right)
 \frac{{\bm{\mathcal Q}}\left({\bm{r}}^\prime,\omega\right)}{\sqrt{\epsilon({\bm{r^\prime}})}}\,.
\end{multline}
\subsection{Photon Green's function}
We introduce the in-plane Green's tensor of the photon ${\bm{\mathcal G}}({\bm{r}},{\bm{r}}^{\prime},\omega)$,
which is defined as the Green's tensor of the Maxwell equation. We have previously shown that simple analytical expressions hold in the case of a QD in a homogeneous medium \cite{parascandolo} or in a planar microcavity \cite{tarel2008}.
In the general case, a compact analytical expression cannot be found. Formally, the  Green's function is defined as :
\begin{eqnarray} \label{greendef}
\bigg[\frac{\omega^2}{c^2}-\bm{\Upsilon}({\bm{r}})\bigg] {\bm{\mathcal G}({\bm{r}},{\bm{r}}^\prime,\omega)}=\delta({\bm{r}}-{\bm{r}}^\prime)
\end{eqnarray}
where $\bm{\Upsilon}({\bm{r}})$ is a time independent, hermitian, linear differential operator that possesses a complete set of eigenfunctions $\{ \Phi_u({\bm{r}})\}$ where $u$ is a continuous index. The set is considered as orthonormal. This differential problem belongs to the class described by Fredholm theory \cite{economou}. It therefore admits a formal solution in terms of the resolvent representation
\begin{eqnarray} \label{green_un}
{\bm{\mathcal G}({\bm{r}},{\bm{r}}^\prime,\omega)}=\int{du\frac{\Phi_u({\bm{r}})\Phi^*_u({\bm{r}}^\prime)}{\frac{\omega_u^2}{c^2}-\frac{\omega^2}{c^2}}}\,.
\end{eqnarray}
Once obtained the Green's function of the photonic structure, the solution of Eq. (\ref{Qprob}), corresponding to an input field ${\bm{\mathcal Q_0}}({\bm{r}},\omega)$ can be written as follows \cite{martin1998}:
\begin{eqnarray} \label{Qequation}
&&{\bm{\mathcal Q}}\left({\bm{r}},\omega\right)=
{\bm{\mathcal Q_0}}\left({\bm{r}},\omega\right)
\\ \nonumber
\\ \nonumber
&&+4\pi \frac{\omega^2}{c^2} \int\int{{d{\bm{r}}^{\prime }d{\bm{r}}^{\prime \prime}{\bm{\mathcal G}({\bm{r}},{\bm{r}}^\prime,\omega)}\frac{\hat{\bm{\chi}}_{QD}\left({\bm{r}}^{\prime },{\bm{r}}^{\prime \prime},\omega\right)}{\sqrt{\epsilon({\bm{r}}^{\prime })}}
{\bm{\mathcal Q}}\left({\bm{r}}^{\prime \prime},\omega\right)}}\nonumber\,,
\end{eqnarray}
The key assumption of our procedure is that one strongly resonant mode exists and is energetically well distinct from any other spectral feature (discrete or continuous) of the structure under investigation. This is the case for all kinds of high-quality nanocavities. Close to resonance $\omega \approx \omega_c$, the following approximation then holds
\begin{multline} \label{Gapprox}
{\bm{\mathcal G}({\bm{r}},{\bm{r}}^\prime,\omega)} \approx \frac{\Phi_0({\bm{r}})\Phi^*_0({\bm{r}}^{\prime })c^2}{2\omega_c(\omega_c-\omega-i\frac{\kappa}{2})}
\\
+\int{du\frac{\Phi_u({\bm{r}})\Phi^*_u({\bm{r}}^\prime)}{2\omega_u(\omega_u-\omega-i\frac{\kappa_u}{2})}}\,.
\end{multline}
A similar expression was used by Sakoda {\em et al.} \cite{sakoda1996} and Hughes {\em et al.} \cite{hughes} . Here, we neglect the longitudinal optical modes, consistently with the exciton optical selection rules that we assume (see below). In compact form we obtain
\begin{eqnarray} \label{Gfinal}
{\bm{\mathcal G}({\bm{r}},{\bm{r}}^\prime,\omega)} \approx \frac{\Phi_0({\bm{r}})\Phi^*_0({\bm{r}}^{\prime })c^2}{2\omega_c(\omega_c-\omega-i\frac{\kappa}{2})}+{\bm g_c}({\bm{r}},{\bm{r}}^{\prime })\,.
\end{eqnarray}
The resonant cavity mode arises as sharp resonance in the energy-dependent density of the eigenmodes. We have characterized this resonance by a damping constant $\kappa$, that models the finite lifetime of the mode. This step is necessary, as we are approximating an everywhere continuous mode spectrum with one discrete mode plus a nonresonant continuum. Formally, this passage can be justified in terms of the \emph{quasi-mode} theory \cite{Scully1990,Savona1998}, by assuming weak coupling between an ideal undamped cavity mode and the vacuum electromagnetic field outside the cavity. In Eq. (\ref{Gfinal}), ${\bm g_c}({\bm{r}},{\bm{r}}^{\prime })$ represents the contribution of all other modes, and is supposed to be small at $\omega \approx \omega_c$. A complete numerical calculation of cavity eigenmodes, like e.g. that carried out in Ref. (\onlinecite{Andreani2004}), can be used to test this assumption. In the following, we will express ${\bm g_c}({\bm{r}},{\bm{r}}^{\prime })$ as the sum of its real and imaginary parts $a({\bm{r}},{\bm{r}}^{\prime })$ and $ib({\bm{r}},{\bm{r}}^{\prime })$. As shown later,
these are responsible -- respectively --
of a shift and a broadening of the QD emission spectrum. More precisely, the term $b({\bm{r}},{\bm{r}}^{\prime })$ is responsible of the decay of the excited QD into  the continuum of background electromagnetic modes. This determines the free decay rate of the QD, usually denoted as $\gamma$ in CQED.
\section{CQED limit}
As a simple test of our formalism, we can recover the limit of one two-level emitter in a resonant cavity, namely the simplest CQED system. Our derivation has the advantage of relating all CQED parameters to microscopic expressions for the semiconductor QD - nanocavity system under investigation.
\subsection{QD susceptibility tensor}
In semiconductors with cubic symmetry (e.g. InGaAs), the QD susceptibility tensor is expressed as
\begin{equation} \label{Chi}
\hat{\bm{\chi}}_{QD}\left({\bf r},{\bf r}^\prime,\omega\right)=
\frac{\mu_{cv}^2}{\hbar}
\Psi^{ }\left({\bf r}\right)
\Psi^*\left({\bf r^\prime}\right)
\bm{\chi}_{QD}\left(\omega\right)
\left(
\begin{array}{ccc}
1 & 0 & 0 \\
0 & 1 & 0 \\
0 & 0 & 0
\end{array}
\right)\,,
\end{equation}
where $\mu_{cv}$ is the Bloch part of the interband dipole matrix element and $\Psi({\bf r})$ is
the electron-hole wave function in the QD, taken at ${\bf r}={\bf r}_e={\bf r}_h$. Here, we are assuming heavy holes only, hence the $z$-component is uncoupled to the electromagnetic field. We further assume to deal with a single QD transition, having one specific polarization (e.g. along ${\bf x}$). Then, the susceptibility tensor is replaced by a scalar, where
\begin{equation}\label{chisimple}
\bm{\chi}_{QD}\left(\omega\right)=\frac{1}
{\omega_0-\omega-i\frac{\gamma_0}{2}}\,.
\end{equation}
Here, $\gamma_0$ is an additional non-radiative damping rate of the bare QD resonance.
Given the small size of the QD with respect to the cavity mode spatial extension, we can safely approximate $\Psi^{ }\left({\bf r}\right)= \delta({\bf r}-{\bf r_0})$. Then, Eqs. (\ref{Qequation}) and (\ref{Chi}) result in
\begin{multline}\label{Qfinal2}
{\bm{\mathcal Q}}\left({\bm{r}},\omega\right)={\bm{\mathcal Q_0}}\left({\bm{r}},\omega\right)+\\
\quad {\bm{\mathcal M}}\left(\omega\right)\left[\! \frac{\Phi_0({\bm{r}})\Phi^*_0({\bm{r_0}})c^2}{2\omega_c(\omega_{c}-\omega-i\frac{\kappa}{2})}+{\bm g_c}({\bm{r_0}},{\bm{r_0}})\right]{\bm{\mathcal Q}}\left({\bm{r_0}},\omega\right)\,,
\end{multline}
with
\begin{equation}\label{momega}
{\bm{\mathcal M}}\left(\omega\right)=\frac{4\pi\mu_{cv}^2}{\hbar c^2}\frac{\omega^2}{(\omega_0-\omega-i\frac{\gamma_0}{2})\sqrt{\epsilon_M}}\,,
\end{equation}
{\small
where $\epsilon_M=\epsilon({\bm{r_0}})$. This expression is the starting point for computing the spectral properties of the cavity-QD system. It gives direct access to the linear response spectrum of the system. From this, the emission spectrum can also be modeled.

\subsection{Emission spectrum}
We first take all fields at position ${\bf r}_0$. Then, Eq. (\ref{Qfinal2}) can be rewritten in compact form
\begin{eqnarray}\label{rabi}
{\bm{\mathcal Q}}\left({\bm{r_0}},\omega\right)((\tilde{\omega_0}-\omega-i\frac{\gamma}{2})(\omega_c-\omega-i\frac{\kappa}{2})-g^2)
\\ \nonumber
={\bm{\mathcal Q_0}}\left({\bm{r_0}},\omega\right) (\omega_0-\omega-i\frac{\gamma_0}{2})(\omega_c-\omega-i\frac{\kappa}{2})\,,
\end{eqnarray}
with
\begin{eqnarray}\label{rabiparam}
        \tilde{\omega_0} & = & \omega_0-\frac{4\pi\mu_{cv}^2\omega_c^2a({\bm{r_0}},{\bm{r_0}})}{\hbar\epsilon_M c^2}\\
 \gamma &=&\gamma_0+\gamma_r \\
 \gamma_r & = & \frac{8\pi\mu_{cv}^2\omega_c^2b({\bm{r_0}},{\bm{r_0}})}{\hbar\epsilon_M c^2}\\
        g^2&=&\frac{2\pi\mu_{cv}^2\omega_c |\Phi_0({\bm{r_0}})|^2}{\hbar\epsilon_M}
\end{eqnarray}
We then compute the emission spectrum of the system from the linear response equation (\ref{Qfinal2}),
using the {\em virtual oscillating dipole} method \cite{sakoda1997}. The method is based on the assumption that
spontaneous emission is the linear response of the system to vacuum field fluctuations. We therefore solve Eq. (\ref{Qfinal2})
with ${\bm{\mathcal Q_0}}\left({\bm{r}}\right)$ given by the field produced by an oscillating dipole at the QD position,
in the photonic structure. We define:
\begin{equation}
{\bm{\mathcal S_q}}\left(\omega\right)=\frac{\sqrt{4g^2-\frac{(\gamma-\kappa)^2}{4}}(\omega_c-\omega-i\frac{\kappa}{2})}{(\omega_0-\omega-i\frac{\gamma}{2})(\omega_c-\omega-i\frac{\kappa}{2})-g^2}
\end{equation}
Using equation (\ref{rabi}), we find:
\begin{multline}\label{qint}
{\bm{\mathcal Q}}\left({\bm{r}},\omega\right)={\bm{\mathcal Q_0}}\left({\bm{r}},\omega\right)+\\
\quad \frac{4\pi\omega^2\mu_{cv}^2}{ c^2 \hbar \sqrt{\epsilon_M}}\frac{{\bm{\mathcal S_q}}\left(\omega\right)}{\sqrt{4g^2-\frac{(\gamma-\kappa)^2}{4}}}
\left[ \frac{\Phi_0({\bm{r}})\Phi^*_0({\bm{r_0}})c^2}{2\omega_c(\omega_c-\omega-i\frac{\kappa}{2})}+{\bm g_c}({\bm{r_0}},{\bm{r_0}}) \right]{\bm{\mathcal Q_0}}\left({\bm{r_0}},\omega\right)
\,.
\end{multline}
The input field ${\bm{\mathcal Q_0}}\left({\bm{r}},\omega\right)$ in our formalism is the field present in the photonic structure in the absence of the QD. This field can be computed by a Green's function procedure similar to the one presented above. In this case, following the method of Ref \cite{martin1998}, the background dielectric system is the free space, while the perturbation is the photonic structure itself. This procedure is presented in Appendix A. We further neglect the first and third term on the right-hand side of Eq. \ref{qint}, as they are off-resonant with respect to the cavity mode. We obtain the following expression for the emitted field
\begin{multline}\label{Qtotal}
\sqrt{\epsilon({\bm{r}})}\bm{\mathcal E}(\bm{r},\omega)=\\
\quad \frac{4\pi\omega^2\mu_{cv}^2}{ c^2 \hbar }\frac{{\bm{\mathcal S_q}}\left(\omega\right)}{\sqrt{4g^2-\frac{(\gamma-\kappa)^2}{4}}}
\left[ \frac{\Phi_0({\bm{r}})\Phi^*_0({\bm{r_0}})c^2}{2\omega_c(\omega_c-\omega-i\frac{\kappa}{2})} \right]{\bm{\mathcal E_0}}\left({\bm{r_0}},\omega\right)\,.
\end{multline}
This result can be now traced back to the well known expressions for the QD-cavity emission spectrum \cite{andreani1999,sakodabisphere}. We use the fact that ${\bm{\mathcal Q_0}}\left({\bm{r}},\omega\right)$ is in general smoothly varying as a function of $\omega$, as discussed in Appendix A. Hence, we assume ${\bm{\mathcal E_0}}\left({\bm{r_0}},\omega\right)\approx \bm{\mathcal E_0}$. Then
\begin{equation}\label{final_result}
\left|\frac{{\bm{\mathcal E}}\left({\bm{r}},\omega\right)}{{\bm{\mathcal E_0}}}\right|^2
={\bm{\mathcal F}}\left({\bm{r}},\omega\right){\bm{\mathcal S}}\left(\omega\right)\,,
\end{equation}
with the semiconductor cavity form factor expressed as
\begin{equation}\label{formfact}
{\bm{\mathcal F}}\left({\bm{r}},\omega\right)=\left|   \frac{4\pi\omega^2\mu_{cv}^2}{ c^2 \hbar \sqrt{4g^2-\frac{(\gamma-\kappa)^2}{4}}}
\left[ \frac{\Phi_0({\bm{r}})\Phi^*_0({\bm{r_0}})c^2}{2\omega_c(\omega_c-\omega-i\frac{\kappa}{2}) \epsilon({\bm{r}})} \right]\right|^2\,.
\end{equation}
The remaining factor in Eq. (\ref{final_result}) is the emission spectrum of the QD, as in atomic CQED, ${\bm{\mathcal S}}\left(\omega\right)=|{\bm{\mathcal S_q}}\left(\omega\right)|^2$, expressed in the resonant case ($\omega_c=\omega_0$) as
\begin{equation}
{\bm{\mathcal S}}\left(\omega\right)=\left|\frac{\Omega_+-\omega_0+i\frac{\kappa}{2}}{\omega-\Omega_+}-\frac{\Omega_--\omega_0+i\frac{\kappa}{2}}{\omega-\Omega_-}\right|^2\,,
\end{equation}
with
\begin{equation}
\Omega_\pm=\omega_0-\frac{i}{4}(\gamma+\kappa) _\pm \sqrt{g^2-\left(\frac{\gamma-\kappa}{4}\right)^2}\,.
\end{equation}

This is the usual CQED result \cite{carmichael, andreani1999}. We see from equation (\ref{rabiparam}) that the coupling of the QD to the electromagnetic field of the modes other than the cavity mode, produces a radiative shift and an additional radiative damping, respectively proportional to the real and imaginary parts $a({\bm{r_0}},{\bm{r_0}})$ and $b({\bm{r_0}},{\bm{r_0}})$ of the photon Green's function. The shift simply redefines the resonant frequency and will be neglected in the following.
The background electromagnetic field however, has also an impact on the radiative damping of the QD. In particular,
$\gamma_r=\frac{4\pi\mu_{cv}^2\omega_c^2 b({\bm{r_0}},{\bm{r_0}})}{\hbar\epsilon_M c^2}$, originating from the term $b({\bm{r_0}},{\bm{r_0}})$, models the radiative decay into the nonresonant background electromagnetic field. Within the CQED formalism, the QD decay rate denoted by $\gamma$ can be linked to our result by defining the damping rate $\gamma=\gamma_0+\gamma_r$.

For example, we use Eq. (\ref{rabiparam}) to compute the Rabi splitting of a semiconductor QD embedded in a photonic crystal nanocavity. We model the optical cavity mode as a Gauss-shaped mode $\Phi_0(\bm{r})$ with spatial extension corresponding to the typical size of a mode in this system. We further assume zero QD-cavity detuning, and the QD position centered at ${\bm{r}_0=0}$. By defining the volume $V_m$ of the Gauss mode, we find:
\begin{equation}\label{somega}
g^2=\frac{2\pi \mu_{cv}^2 \omega_c}{\hbar\epsilon_MV_m}\,.
\end{equation}
This expression coincides with that obtained in Ref. \onlinecite{andreani1999}. With realistic numerical values for InAs QDs in photonic crystal nanocavities ($\mu_{cv}^2=480meVnm^3$, $V_m=0.04\mu m^3$), we find $g\approx 200 \mu eV$.
\section{Beyond the macroatom picture}
Recent studies have demonstrated that a semiconductor QD displays spectral features that depart from the simple picture of a two-level system. One typical example is the non-perturbative coupling to acoustic phonons, resulting in broad phonon sidebands in the exciton spectrum. This mechanism has now been extensively characterized both theoretically \cite{zimmermann,krumm,muljarov} and experimentally \cite{vasanelli,borri,Besombes}. Another mechanism that has been recently investigated is the transition between multi-exciton manifolds, involving the continuum of excited states of each manifold (sometimes referred to as ``shakeup process''). This mechanism has proven very effective especially when a QD is embedded in a resonant cavity, giving rise to intense PL at the cavity mode even at very large cavity-QD detuning -- the {\em cavity feeding} mechanism. The formalism discussed here can be generalized to situations like the first one, characterized by a homogeneous spectral modification, by replacing the simple QD susceptibility (\ref{chisimple}) with the appropriate model. Here, as an example, we discuss the case of exciton-acoustic phonon coupling with formation of phonon sidebands. We are still interested in determining the emission spectrum in the form (\ref{final_result}).
\subsection{QD susceptibility tensor}
The coupling of one exciton to the LA-phonon band is described exactly, through the solution of the {\em independent Boson model} \cite{mahan,zimmermann}. It has however been shown \cite{krumm} that a very good account of the exciton spectrum can be obtained already at the 2nd Born perturbation level, with the advantage of having a simple expression for the exciton-phonon self-energy. We thus rewrite the QD exciton susceptibility including the exciton-phonon self energy as
\begin{equation} \label{Chiph}
\bm{\chi}_{QD}\left(\omega\right)=\frac{1}{\omega_0-\omega-i\frac{\gamma_0}{2}+\Sigma(\omega)}\,,
\end{equation}
where, within second Born approximation and restricting to only one phonon band,
\begin{equation*}
\Sigma(\omega)=\sum_{q}\left[
\frac{{\mid g_{q}^x\mid}^2(1+n(q))}{\omega+i\frac{\gamma}{2}-\omega_0-\omega(q)}
+
\frac{{\mid g_{q}^x \mid}^2(n(q))}{\omega+i\frac{\gamma}{2}-\omega_0+\omega(q)}\right]\,.
\end{equation*}
Here, $n(q)$ is the Bose-Einstein equilibrium phonon occupation at temperature $k_BT$. We consider the case of deformation potential coupling with acoustic phonons of dispersion $\omega_{q}=qs$ ($q=\mid {\bm{q}} \mid$), where $s$ is the sound velocity, as in Ref. \onlinecite{zimmermann}. In Fig. \ref{fig3} we display the imaginary part of the QD susceptibility, as computed at $k_BT=10K$ for an InAs QD of 10nm diameter. It should be noted that the phonon spectral features do not depend specifically on the shape of the exciton wave function but are only determined by its volume \cite{zimmermann}. In the plot, we notice the pronounced sidebands compared to the spectrum of an ideal exciton. The sidebands are more pronounced on the high energy side, where they are determined by acoustic phonon emission.
\begin{figure}[ht]
 \centerline{\includegraphics*[width=1\linewidth]{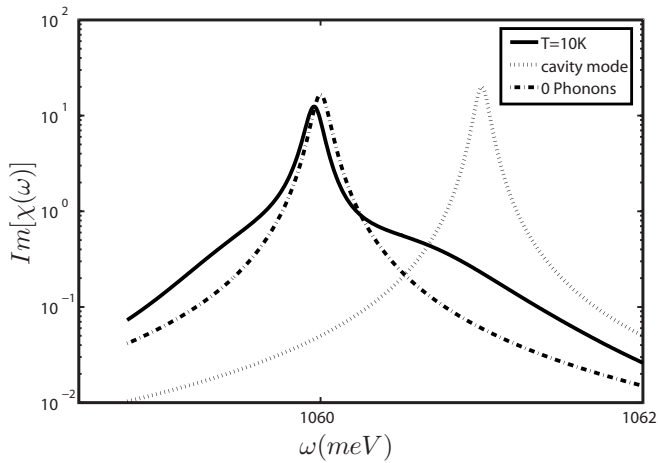}}
  \caption{Imaginary part of the quantum dot susceptibility in presence of LA-phonon coupling (full) and without phonons (dashed), computed at T=10K. As an illustration, we plot the cavity mode optical density at 1 meV positive detuning (dotted).}
  \label{fig3}
\end{figure}
\subsection{Emission spectrum}
Intuitively, the emission intensity at the cavity-mode frequency depends on the optical density of the underlying exciton spectrum. Hence, the presence of acoustic phonon sidebands is expected to enhance this PL intensity, when the cavity is detuned from the exciton. This is illustrated in Fig. \ref{fig3}, where the cavity mode spectrum is plotted at 1 meV positive detuning with respect to the exciton peak.

We use the QD susceptibility (\ref{Chiph}) to compute the emission spectrum (\ref{final_result}). The form factor ${\mathcal F}({\bm r},\omega)$ is still expressed as (\ref{formfact}), while the QD emission spectrum now reads
\begin{equation}\label{spectreph}
{\bm{\mathcal S}}\left(\omega\right)=\Big|\frac{\sqrt{4g^2-\frac{(\gamma-\kappa)^2}{4}}(\omega_c-\omega-i\frac{\kappa}{2})}{(\omega_0-\omega-i\frac{\gamma}{2}+\Sigma(\omega))(\omega_c-\omega-i\frac{\kappa}{2})-g^2}\Big|^{2}\,.
\end{equation}
As expected, the exciton-phonon coupling results in a modified emission spectrum. In particular, the exciton-phonon self energy is responsible for a modified intensity at the cavity mode frequency and a small polaron shift of the exciton frequency. In Fig. \ref{fig4}, we plot the computed spectrum at $k_BT=10$ K for various values of the exciton-cavity detuning. While the strong coupling features remain essentially unchanged for zero-detuning
(see panel b), we can clearly see in panels (a), (c), and (d) that phonon sidebands can efficiently emit through the cavity mode, as also found by other theoretical approaches \cite{Auffeves09,Tawara09}. As a consequence, the peak at frequency $\omega \approx \omega_c$ is enhanced with respect to the simple CQED model, provided the detuning is not larger than the energy extent of the phonon bands. This effect was widely investigated experimentally during the last years. It has been reported for QDs embedded in various systems such as photonic crystal nanocavities \cite{Hennessy,Kaniber08,Ota09} and micropillars \cite{suffczynski2009}. We can see in Fig. \ref{fig4}(d) that for detuning exceeding the typical broadband width, this feature starts to disappear, as at large detuning the sideband essentially vanishes. Moreover, the strong asymmetry in $\hat{\bm{\chi}}_{QD}\left(\omega\right)$ at low temperatures has for consequence that the persistence of the peak is small for negative detuning (see panel (a)). Finally, in Fig. \ref{fig5}, we show the influence of the temperature on the emission spectrum $S(\omega)$. The phonon sidebands grow with
temperature and result in an increased emission through the cavity mode.
\begin{figure}[ht]
  \centerline{\includegraphics*[width=1\linewidth]{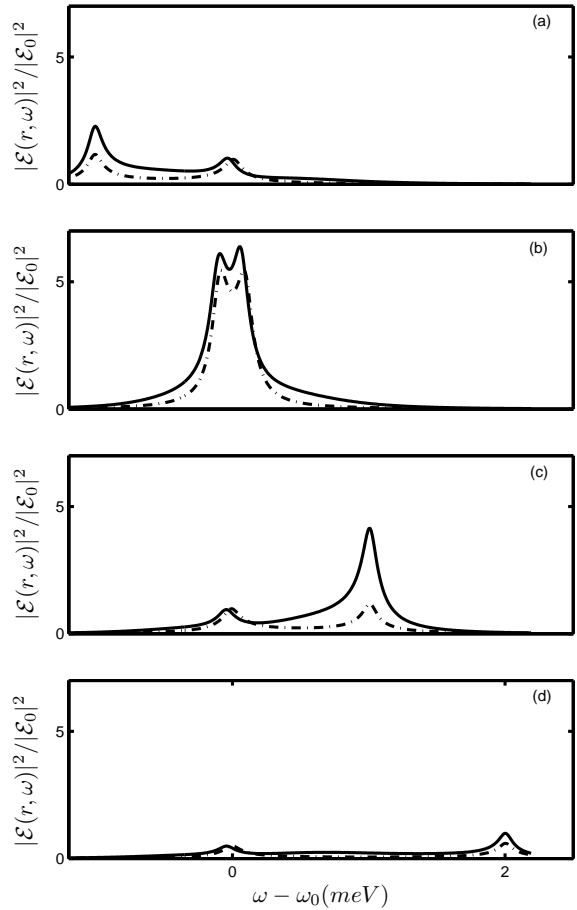}}
  \caption{Plot of the emission spectrum of the QD-cavity system in the presence of phonons (solid line) for different detunings $D=\omega_{0}-\omega_C$. a) $D=-1 meV$, b) $D=0 meV$, c) $D=1 meV$, d) $D=2 meV$ ($k_BT$=10K). Comparison with no phonons (dashed line) is also given.}
  \label{fig4}
\end{figure}
\begin{figure}[ht]
 \centerline{\includegraphics*[width=1.2\linewidth]{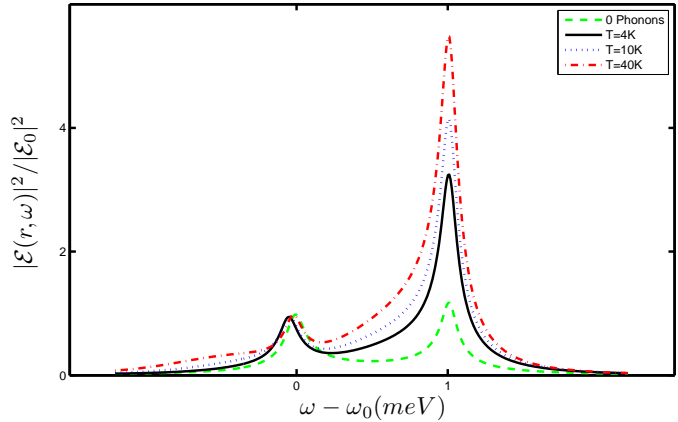}}
  \caption{Emission spectrum of the QD-cavity system plotted for different values of $k_BT$ (arbitrary units), $\omega_c-\omega_{0}=1 \ meV$.}
  \label{fig5}
\end{figure}

This result suggests that the enhanced emission at the cavity mode, observed in several recent experiments, could be attributed to the phonon sideband mechanism. We point out however, that this cavity feeding phenomenon has been observed also when the exciton-cavity detuning is much larger than the typical width of the phonon broadbands, ie a few meV. These observations are accompanied by a superlinear dependence of the cavity mode PL on the excitation power. The phonon sideband mechanism, on the other hand, is expected to provide a spectrum that depends linearly on the excitation power. The phonon sideband model is thus expected to hold mostly at small detuning. The observations of cavity feeding at larger detuning are most likely due to multi-exciton emission, partially involving wetting layer states, as has been recently discussed \cite{Winger2009,Chauvin09,Hughes09,Kaniber08}.

\subsection{Influence of neighboring QDs}

One major assumption of this model is that there is only one QD dot located within the region in which the cavity mode is extending. Given the density of the QD ensemble and the spatial extension of the nanocavity mode, it might well be that \emph{spectator} QDs -- i.e. additional QDs present in the cavity -- contribute to the emission spectrum. These QDs are most likely weakly coupled to the cavity mode because of strong energy detuning or of smaller spatial overlap with the mode wave function. It has been recently suggested \cite{LaussyPRL,LaussyPRB} that, if these QDs are excited in addition to the main QD, the resulting emission spectrum is substantially modified, sometimes even leading to a recovery of strong coupling in a situation that would be of weak coupling if only the main QD was excited. In Refs. \onlinecite{LaussyPRL} and \onlinecite{LaussyPRB}, this effect has been modeled by an additional pump term acting on the cavity mode. Here, we can account for the presence of additional QDs in a natural way, by generalizing the expression for the single-QD susceptibility (\ref{Chi}). The new susceptibility then reads
\begin{eqnarray}
\lefteqn{\hat{\bm{\chi}}_{QD}\left({\bf r},{\bf r}^\prime,\omega\right)=
\frac{\mu_{cv}^2}{\hbar}
\left(
\begin{array}{ccc}
1 & 0 & 0 \\
0 & 1 & 0 \\
0 & 0 & 0
\end{array}
\right)}\nonumber\\
&\times&\left[
\Psi^{ }\left({\bf r}\right)
\Psi^*\left({\bf r^\prime}\right)
\bm{\chi}_{QD}\left(\omega\right)
+\sum_j\Psi_j^{ }\left({\bf r}\right)
\Psi_j^*\left({\bf r^\prime}\right)
\bm{\chi}_j\left(\omega\right)\right]\,,
\label{Chimultiple}
\end{eqnarray}
where
\begin{equation}\label{chim}
{\bm{\chi}_j}\left(\omega\right)=\frac{1}
{\displaystyle \omega_j-\omega-i\frac{\gamma_{0j}}{2}+\Sigma_j(\omega)}\,.
\end{equation}
Here, the $j$-th QD has parameters defined analogously to those of the main QD. Starting from this expression, the derivation of the emission spectrum can be carried out analogously to the single-QD case. In particular, when determining the input field ${\bm{\mathcal Q_0}}\left({\bm{r}},\omega\right)$ as described in Appendix A, the virtual oscillating dipole will consist of a sum of terms originating from the different QDs, with relative weights $B_j$ that express the contribution of each QD to the initial state of the emission process. It should be pointed out however, that the present model is based on the linear response to the Maxwell field. In Ref. \onlinecite{LaussyPRL} the additional pump term enhances the strong coupling by compensating for the cavity losses -- a mechanism that can be traced back to the gain produced by the excitation of the additional QDs in the cavity. This enhancement cannot be reproduced by our model, as it would require accounting for nonlinear optical response. Eq. (\ref{Chimultiple}) thus can model the presence of spectator QDs only in the limit of very small average population of each QD, for which the linear assumption holds.

\section{Conclusion}
In conclusion, we have shown that the Greens function formalism is a powerful tool to relate quantitatively the usual atom-CQED parameters to the description of any QD-cavity system. We also extended this formalism to a QD weakly coupled to LA phonons. Thus we underlined that the difference of a QD to a simple two level system is greatly enhanced when the quantum dot is placed inside a nanocavity. It makes possible that a PL peak of considerable amplitude remains at cavity frequency even for large detuning compared to the Rabi splitting, but limited to a few $mev$. This shows that the broad spectral features provided by the environment of a QD play a key role in the cavity-QD systems response.

\pagebreak


\appendix
\section{Determination of the bare cavity electric field}

To determine ${\bm{\mathcal Q_0}}({\bm{r}},\omega)$, we will follow the approach proposed in Ref. \onlinecite{martin1998}. We have

\begin{equation}
 \bm{\mathcal Q_0}=\sqrt{\epsilon({\bm{r}})}\bm{\mathcal E_0}(\bm{r},\omega)
\end{equation}
with

\begin{eqnarray}
&&\bm{\nabla}\wedge\bm{\nabla}\wedge
{\bm{\mathcal E_0}}\left({\bm{r}},\omega\right)-\frac{\omega^2}{c^2}
\epsilon({\bm{r}}) {\bm{\mathcal E_0}}\left({\bm{r}},\omega\right)
=0
\nonumber
\end{eqnarray}
That is, with $\Delta \epsilon({\bm{r}})=\epsilon({\bm{r}})-\epsilon_B$,

\begin{eqnarray}
&&\bm{\nabla}\wedge\bm{\nabla}\wedge
{\bm{\mathcal E_0}}\left({\bm{r}},\omega\right)-\frac{\omega^2}{c^2}
\epsilon_B {\bm{\mathcal E_0}}\left({\bm{r}},\omega\right)=\frac{\omega^2}{c^2}
\Delta\epsilon({\bm{r}}) {\bm{\mathcal E_0}}\left({\bm{r}},\omega\right)
\nonumber
\end{eqnarray}
We define $\bm{\mathcal E_B}(\bm{r},\omega)$ as a solution of
\begin{eqnarray}
&&\bm{\nabla}\wedge\bm{\nabla}\wedge
{\bm{\mathcal E_B}}\left({\bm{r}},\omega\right)-\frac{\omega^2}{c^2}
\epsilon_B {\bm{\mathcal E_B}}\left({\bm{r}},\omega\right)=0
\,,\nonumber
\end{eqnarray}
which is indeed a plane wave. At this point, we will use the using the {\em virtual oscillating dipole} method \cite{sakoda1996}.
We replace ${\bm{\mathcal E_B}}$ by a point source centered in ${\bm{r_0}}$: ${\bm{\mathcal E_B}}\left({\bm{r}},\omega\right)=B\delta({\bm{r}}-{\bm{r_0}})$.
Using the background Green's function defined as:
\begin{eqnarray}
&&\bm{\nabla}\wedge\bm{\nabla}\wedge
{\bm{\mathcal G_B}}\left({\bm{r}},\omega\right)-\frac{\omega^2}{c^2}
\epsilon_B {\bm{\mathcal G_B}}\left({\bm{r}},\omega\right)=\delta({\bm{r}}-{\bm{r}}^\prime)
\,,\nonumber
\end{eqnarray}
we have:
\begin{eqnarray}
&&{\bm{\mathcal E_0}}\left({\bm{r}},\omega\right)=
{\bm{\mathcal E_B}}\left({\bm{r}},\omega\right)
\\
&&+\int_V{{d{\bm{r}}^{\prime }{\bm{\mathcal G_B}({\bm{r}},{\bm{r}}^\prime,\omega)}\frac{\omega^2}{c^2}\Delta\epsilon({\bm{r}}^{\prime })}}
{\bm{\mathcal E_B}}\left({\bm{r}}^{\prime },\omega\right)\\
&&=B\left[\delta({\bm{r}}-{\bm{r_0}})+{\bm{\mathcal G_B}({\bm{r}},{\bm{r_0}},\omega)}\frac{\omega^2}{c^2}
\Delta\epsilon({\bm{r_0}})\right]
\,.
\end{eqnarray}
In this expression, ${\bm{\mathcal G_B}({\bm{r}},{\bm{r}}^\prime,\omega)}$ is a slowly varying and especially non-resonant function
of $\omega$. Then so does ${\bm{\mathcal E_0}}\left({\bm{r}},\omega\right)$ and finally:
\begin{equation} \label{Max}
 \bm{\mathcal Q_0}\left({\bm{r}},\omega\right)=\sqrt{\epsilon({\bm{r}})}\bm{\mathcal E_0}(\bm{r},\omega)\,,
\end{equation}
with $ \bm{\mathcal Q_0}\left({\bm{r}},\omega\right)$ a function with no resonance.

\end{document}